# Energy-spread measurement of triple-pulse electron beams based on the magnetic dispersion principle

Yi Wang, Qin Li, Zhiyong Yang, Huang Zhang, Hengsong Ding, Anmin Yang, Minhong Wang

Key Laboratory of Pulsed Power, Institute of Fluid Physics, China Academy of Engineering Physics, Mianyang 621900, Sichuan Province, China

**Abstract:** The energy-spread of the triple-pulse electron beam generated by the Dragon-II linear induction accelerator is measured using the method of energy dispersion in the magnetic field. A sector magnet is applied for energy analyzing of the electron beam, which has a bending radius of 300 mm and a deflection angle of 90°. For each pulse, both the time-resolved and the integral images of the electron position at the output port of the bending beam line are recorded by a streak camera and a CCD camera, respectively. Experimental results demonstrate an energy-spread of less than ±2.0% for the electron pulses. The cavity voltage waveforms obtained by different detectors are also analyzed for comparison.

## 1. Introduction

The linear induction accelerator (LIA) is able to generate charged particle beam pulses with extremely high currents, which has been well developed and widely applied in various fields since it was first proposed in 1950s, such as flash radiography [1], high power microwave generation [2] and heavy ion fusion [3,4]. In order to investigate the hydrodynamic process of high explosives, electron beam pulses are generated and accelerated to ~MeV energy in the LIA and finally focused onto a high-Z convertor to produce X-ray photons through the bremsstrahlung radiation [5]. The temporal width of the pulse is typically tens of nanoseconds, which is capable of recording an inner stopped-motion image of a dense object with a relatively small motion blur. In order to obtain fine details of the acquired image, it is strongly demanded to reduce the X-ray source to a spot size as small as possible, which is quoted as the evaluation of the resolving ability.

For the ideal condition, the electron beam can be focused to an infinitesimally small spot, which is expected to generate a point X-ray source. However, the reduction of the X-ray spot size is limited by many factors, including the space charge effect, the spherical aberration of the lens, the beam emittance, the energy-spread, etc. [6] The effect of the electron energy spread on the spot size comes from two aspects. On the one hand, the minimal radius of the spot size determined by the dispersion aberration of the focusing lens is directly dependent on the energy-spread of the electron beam. [7] On the other hand, the energy spread of the electron beam will aggravate the Corkscrew oscillation induced by the titled beam injection, an inaccurate alignment of the solenoidal field, which finally gives rise to a bigger spot size and a larger spatial jitter. [8] The energy-spread of the electron beam is mainly determined by the waveform and the synchronization of the induction accelerating voltages as well as the load effect of the intense-current beam. [9] In this paper, the method based on magnetic dispersion is used to measure the energy-spread of triple-pulse electron beams produced by the Dragon-II LIA. Besides, the cavity voltage waveforms detected by both the capacitor voltage probe (CVP) and resistor voltage divider (RVD) are analyzed for comparison.

## 2. Principle of magnetic dispersion

In the magnetic field, electrons with different energies move along distinct bending orbits. Such a spatial dispersion of charged particles results from the effect of Lorentz force, which serves as the centripetal force of the vertical deflection and follows the relation of [10]

$$\frac{mv^2}{\rho} = evB,\quad(1)$$

where $m$ is the particle mass, $v$ is the scalar velocity, $\rho$ is the bending radius of the orbit, $e$ is the electron charge, and $B$ is magnetic induction intensity perpendicular to velocity. For a relativistic electron, the momentum $p = mv$ can be further expressed as

$$p = \beta\gamma m_0 c,\quad(2)$$

where $\beta = v/c$ with c being the light velocity in vaccum, $\gamma = (1-\beta^2)^{-1/2}$ is the relativistic factor, and $m_0$ is the rest mass of electron. The kinetic energy $E_k$ is given by

$$E_k = (\gamma - 1)m_0 c^2.\quad(3)$$

Substituting Eqs. (1) and (2) into (3) yields the following relation between the electron kinetic energy and the orbit curvature radius as

$$E_k = m_0 c^2 \left[\sqrt{1+\left(\frac{eB\rho}{m_0 c}\right)^2} - 1\right].\quad(4)$$

For a sector magnet with uniform magnetic field and normal entry and exit, the radial coordinates, $x_1$, $x_2$ and the slopes, $x_1'$, $x_2'$ of a trajectory at two different positions are linked with the momentum deviation $\Delta p/p_0$ by the transfer matrices [11], i.e.

$$\begin{pmatrix} x_2 \\ x_2' \\ \frac{\Delta p}{p_0} \end{pmatrix} = \begin{pmatrix} 1 & L_2 & 0 \\ 0 & 1 & 0 \\ 0 & 0 & 1 \end{pmatrix} \begin{pmatrix} \cos\varphi & \rho_0 \sin\varphi & \rho_0(1-\cos\varphi) \\ -\frac{1}{\rho_0}\sin\varphi & \cos\varphi & \sin\varphi \\ 0 & 0 & 1 \end{pmatrix} \begin{pmatrix} 1 & L_1 & 0 \\ 0 & 1 & 0 \\ 0 & 0 & 1 \end{pmatrix} \begin{pmatrix} x_1 \\ x_1' \\ \frac{\Delta p}{p_0} \end{pmatrix},\quad(5)$$

where $L_1$ and $L_2$ are the objective distance and the image distance, respectively, $\varphi$ is the deflection angle, $\rho_0$ is the bending radius of the central ray inside the magnetic field. Consider the symmetric drift-space of $L_1 = L_2 = \rho_0$ and the bending angle of $\varphi = 90°$, and set the radial coordinate of the object position to be $x_1 = 0$, the relation between the momentum difference and the parameters of the trajectory can be simplified as

$$\frac{\Delta p}{p} = \frac{x_2}{2\rho_0},\quad(6)$$

Then the corresponding difference in kinetic energy is given by

$$\frac{\Delta E_k}{E_{k0}} = \frac{\gamma+1}{\gamma}\frac{x_2}{2\rho_0},\quad(7)$$

where $E_{k0}$ is the kinetic energy of the electron moving on the reference trajectory ($\rho = \rho_0$). It is seen that the radial coordinate of the trajectory at the image plane is only dependent on the kinetic energy but not the slope at the object position. Figure 1 shows the numerically calculated orbits of the electrons, which have distinct image positions for different kinetic energies (18.5

MeV and 19.5 MeV). The lengths of the drift spaces before and after the magnetic field (B = 2170 Gs) are both 300 mm. The reference orbit inside the magnetic field has a bending radius of 300 mm and a deflection angle of $90°$. The electrons of the same kinetic energy are focused to the same point on the image plane though starting with different azimuth angles. By measuring the image position with respect to the reference trajectory, the electron energy can be obtained and the energy spread is then calculated by

$$\varepsilon = \pm \frac{E_{k,max} - E_{k,min}}{2E_{k,mean}} \times 100\% \qquad (8)$$

where $E_{k,max}$, $E_{k,min}$ and $E_{k,mean}$ are the maximum, minimum and mean kinetic energies, respectively, corresponding to the maximum, minimum and mean $x_2$.

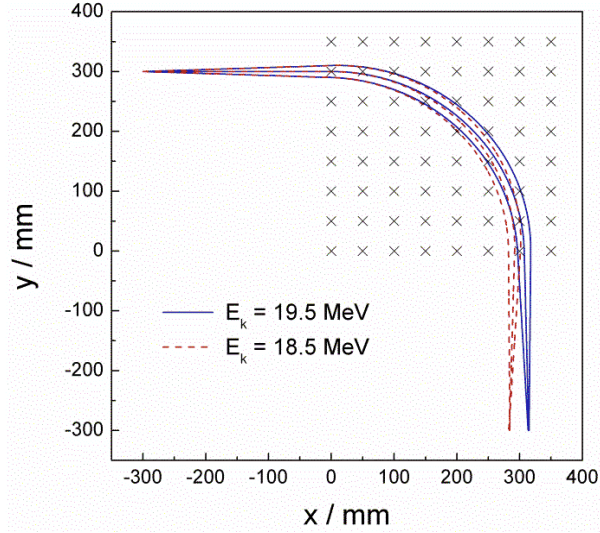

Fig. 1 Energy dispersion of the electron beam in the magnetic field. The magnetic induction intensity is B = 2170 Gs.

## 3. Experimental setup and measurements

As shown in Fig. 2, a graphite slit with the axial length of 80 mm and the slit width of 1 mm is fixed on the input port of the vacuum tube, which restricts the entrance position of electrons exactly on the central trajectory. A sector magnet with the bending radius of 300 mm and the deflection angle of $90°$ is carefully constructed on the beam line for measurement. The magnet has been designed and modified to guarantee the effective edges of the magnetic extended fringing field match with the deflection part of the tube. The lengths of the object-side and the image-side drift spaces are both 300 mm. A quartz glass slice is placed at the output port of the tube in order to generate optical photons through Cherenkov radiation [12] when the electrons exit the tube and impinge on the glass. The optical image is splitted into two paths by a splitting prism and then received by a streak camera and a CCD camera, respectively. The streak camera is able to record a continuous time-resolved image position during a pulse. The CCD camera, which is coupled with a micro channel plank for gating-time control, is used to record the integrated image of a whole pulse. In order to provide a stable uniform magnetic field, a power supply is used to load a constant current on the excitation coil of the magnet. The corresponding relation between the magnetic intensity and the loaded current has been scaled with a Gaussmeter in advance. The stabilization of the current is limited within 10 mA, which denotes a fluctuation of the magnetic field less than 5 Gs during the measurement.

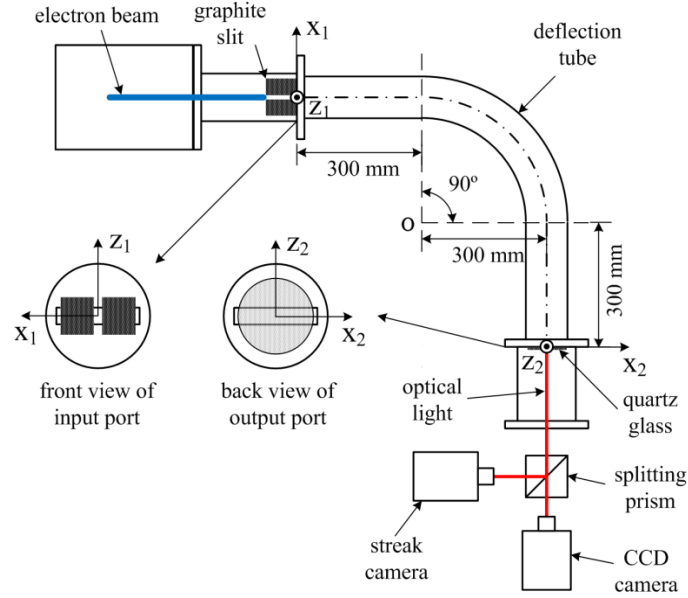

Fig. 2 Experimental setup for energy-spread measurement.

Experiments are carried out on the Dragon-II LIA for the energy-spread measurement of the electron beam. Three electron pulses (denoted as A, B and C, respectively) are generated at a time during one experiment. The time distances between each two neighboring pulses are ~400 ns. For each measurement, the external signal for triggering the cameras is tuned in order to synchronize with an electron pulse to be detected. Then both the time-solved image and the integrated image of the electron at the exit port are recorded, by which the energy spread of the electron beam can be calculated according to the relation between the electron energy and the image position.

### 4. Results and discussions

For each experiment, a triple-pulse electron beam passes through the magnetic analyzer, among which the image of one chosen pulse at the exit port is recorded. The alternation of the image position actually denotes the variation of the electron energy during the pulse. The images captured by the streak camera and the CCD camera in the experimental measurements are shown in Fig. 3. A slit is installed at the entry of the streak camera to further limit the z-axis of the image, the direction of which is perpendicular to that of the graphite slit. So the energy of the electron pulse manifests itself as a continuously time-variable spot position on the streak camera image. The CCD camera records the integral image of the electron at the exit port of the tube, the width of which in the x-axis represents the energy range of a whole pulse. The experimental results are listed in Table 1. We mainly consider the flat top part of the electron beam pulse, which is ~50 ns for pulse A and ~60 ns for pulses B and C. The results reveal that the energy spread of the flat top of the electron beam pulse is not more than ±2.0%. Specifically, pulse A has the smallest energy spread ($\varepsilon = \pm1.35\%$) while pulse C has the largest one ($\varepsilon = \pm1.95\%$).

The energy-spread of the electron beam mostly results from imperfect square voltage waveforms and asynchrony between different accelerating cavities. The triple-pulser of the Dragon-II LIA consists of three square-pulse sections, a confluent/blocking network, and a control/triggering modular unit. [13,14] The pulser has been designed for load-matching, which helps to reduce the reflections of the pulses but still cannot completely eliminate the effect. The reflection of the former voltage pulse(s) will add on the latter one(s). Therefore, the energy spread of a former electron beam pulse is generally smaller than a latter one does.

The cavity voltage waveform is detected by the CVP and the RVD. As shown in Fig. 4, the waveforms obtained by the two probes slightly differ from each other at the flat top part. The detected waveform is closely related to the time difference between the voltage signal and the beam-load signal arriving at the detector. The CVP is located close to the accelerating gap while the RVD is at the outside of the cavity. Thus the CVP is expected to provide a waveform closer to the real one. This is verified by the fact that the CVP waveform accords much better with the result of magnetic analyzing. The current waveform obtained by the beam position monitor (BPM) at the upstream of the tube is also given, which indicates that the camera images correspond to the central part of the electron beam.

The error of the experiment mainly results from an inaccurate value of the magnetic field intensity and a deviation of the electron position. The former one is composed of the accuracy of magnet scaling and the stabilization of the current loaded, which give an error of less than 15 Gs in the magnetic field intensity. The constitution the latter one includes the deviations in positioning the input graphite slit, the deflection tube and the magnet, which correspond to an error of totaling 4 mm in the image position at the output port. The error of the kinetic energy is calculated to be $E_{k,error}$ = 0.27 MeV at ~19 MeV. Because the energy spread is a relative difference of the kinetic energy, the error of the energy spread is mainly determined by the width of the graphite slit and the discrimination of electron position from the image. Then the error of the energy spread is calculated to be $\varepsilon_{error}$ = 0.68% (absolute value) in the experiment.

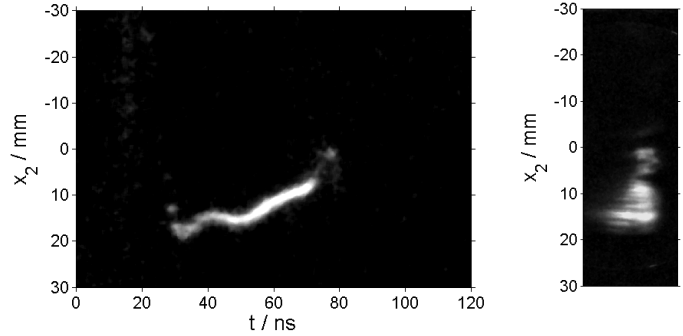

(a)

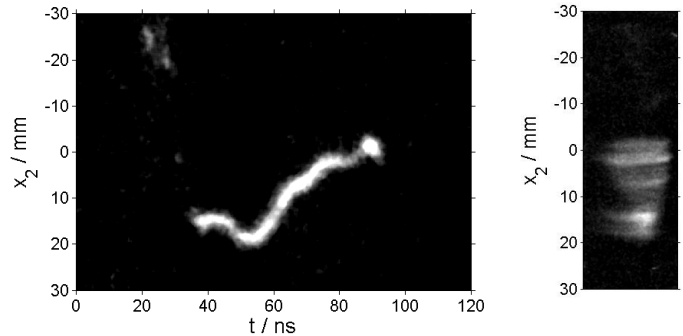

(b)

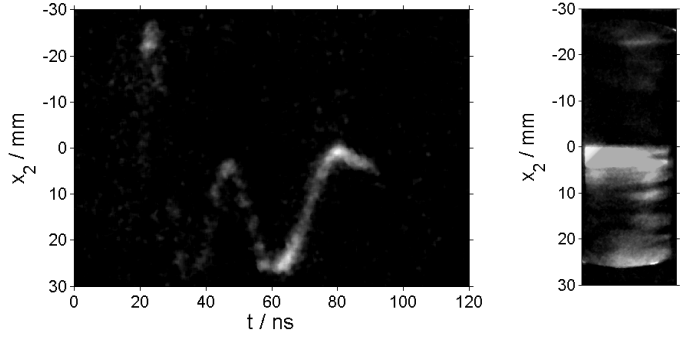

(c)

Fig. 3 Images captured by streak camera (left image) and CCD camera (right image). (a) No. #1, pulse A (B = 2167 Gs, $E_k(x_2=0)$ = 19.0 MeV); (b) No. #2, pulse B (B = 2143 Gs, $E_k(x_2=0)$ = 18.8 MeV); (c) No. #3, pulse C (B = 2143 Gs, $E_k(x_2=0)$ = 18.8 MeV).

Table 1 Experimental results of the energy-spread measurement.

| No. | Pulse | B /Gs | $\tau_{flat}$ /ns | $x_2$ / mm | | | $E_k$ / MeV | | | ε |
|---|---|---|---|---|---|---|---|---|---|---|
| | | | | min | max | mean | min | max | mean | |
| #1 | A | 2167 | 50 | 1.5 | 17.4 | 9.5 | 19.03 | 19.55 | 19.29 | ±1.35% |
| #2 | B | 2143 | 60 | -1.0 | 18.2 | 8.6 | 18.74 | 19.36 | 19.05 | ±1.63% |
| #3 | C | 2143 | 60 | 1.7 | 25.3 | 13.5 | 18.83 | 19.58 | 19.21 | ±1.95% |

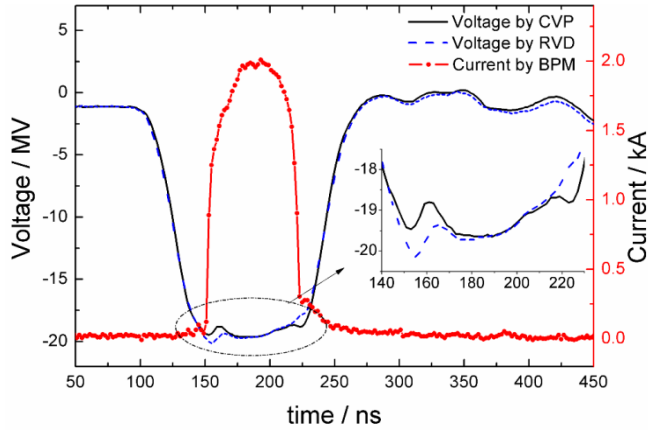

(a)

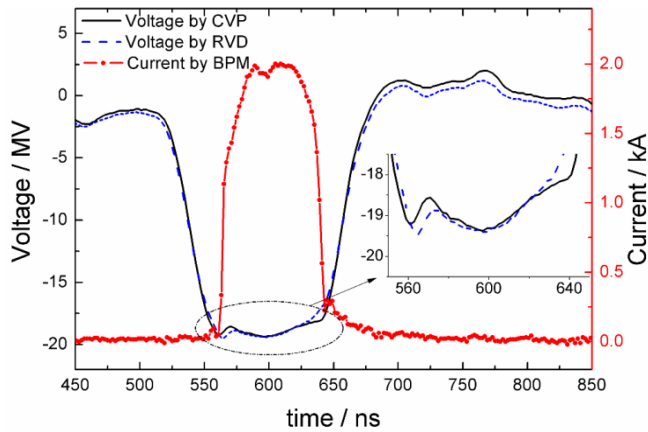

(b)

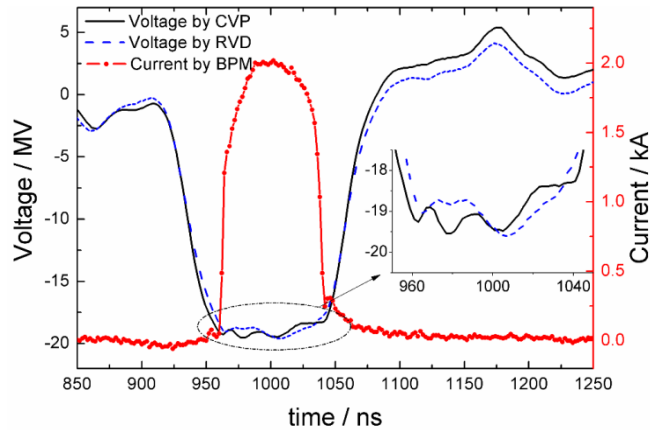

(c)

Fig. 4 Voltage and current waveforms of electrical signals. (a) No. #1, pulse A; (b) No. #2, pulse B; (c) No. #3, pulse C.

## 5. Conclusion

The method of magnetic analyzing is applied to measure the energy-spread of the triple-pulse electron beam produced by the Dragon-II LIA. The beam line for diagnostic is designed to have symmetric drift-space lengths at the upstream and downstream of the vector magnet, which has a bending radius of 300 mm and a deflection angle of 90°. The energy spread of the electron beam for different pulses are all within ±2.0% for the flat-top part. Due to the effect of the cavity voltage reflection, the first pulse of the electron beam has the smallest energy-spread while the last pulse has the largest one. Due to the time difference caused by the cavity structure, the voltage waveforms detected by the CVP and RVD show to be slightly different at the flat top. Since the location of the CVP is very close to the accelerating gap, it is expected to provide an exact cavity voltage. Experimental results display a good agreement between the CVP waveform and the energy spectrum of the magnetic analyzer.